\documentclass[conference]{IEEEtran}
\pdfoutput=1
\IEEEoverridecommandlockouts
\usepackage{cite}
\usepackage{amsmath,amssymb,amsfonts}
\usepackage{algorithmic}
\usepackage{graphicx}
\usepackage{textcomp}
\usepackage{xcolor}
\usepackage{comment}

\usepackage{hyperref}
\usepackage{xcolor}
\hypersetup{
    hidelinks,
}

\usepackage[english]{babel}
\usepackage{blindtext}
\usepackage{enumitem}
\usepackage{algorithm2e}
\usepackage{multirow}
\usepackage{enumitem}
\setlist[itemize]{leftmargin=*}
\usepackage{subcaption}
\usepackage{booktabs}

\usepackage[normalem]{ulem}
\usepackage{color, soul}
\sethlcolor{red}

\def\BibTeX{{\rm B\kern-.05em{\sc i\kern-.025em b}\kern-.08em
    T\kern-.1667em\lower.7ex\hbox{E}\kern-.125emX}}
\begin{document}

\title{Progressive Neural Compression for Adaptive Image Offloading under Timing Constraints}



\author{
\IEEEauthorblockN{Ruiqi Wang, Hanyang Liu, Jiaming Qiu, Moran Xu, Roch Gu\'erin, Chenyang Lu} 
\IEEEauthorblockA{\textit{Department of Computer Science and Engineering} \\
\textit{Washington University in St. Louis}\\
St. Louis, Missouri, USA\\
Email:  \{ruiqi.w, hanyang.liu, qiujiaming, moranxu, guerin, lu\}@wustl.edu}
}


\newcommand {\ie} {{\em i.e., }}
\newcommand {\eg} {{\em e.g., }}
\newcommand {\iid} { i.i.d. }
\newcommand {\etal} {{\em et al.}}
\newcommand {\beq} {\begin{equation}}
\newcommand {\eeq} {\end{equation}}
\newcommand {\bequn} {\begin{equation*}}
\newcommand {\eequn} {\end{equation*}}
\newcommand {\bear} {\begin{eqnarray}}
\newcommand {\eear} {\end{eqnarray}}
\newcommand {\bearun} {\begin{eqnarray*}}
\newcommand {\eearun} {\end{eqnarray*}}
\newcommand {\fig}[1]{Fig.~\ref{#1}}
\newcommand {\Eqref}[1]{Eq.~(\ref{#1})}

\newcommand{\rg}[1]{{\color{red}{[RG: #1]}}}
\newcommand{\ac}[1]{{\color{red}{[AC: #1]}}}
\newcommand{\jq}[1]{{\color{red}{[JQ: #1]}}}
\newcommand{\cl}[1]{{\color{red}{[CL: #1]}}}
\newcommand{\wrq}[1]{\textcolor{red}{{[RW: #1]}}}
\newcommand{\blue}[1]{\textcolor{blue}{{#1}}}
\newcommand{\citehere}{\colorbox{pink}{[TODO: Reference]}}
\newcommand{\todo}[1]{\hl{[TODO: #1]}}
\newcommand{\PNC}{PNC}
\newcommand{\new}[1]{{\color{red}{#1}}}
\newcommand{\newrtss}[1]{{\color{orange}{#1}}}

\maketitle
\thispagestyle{plain}
\pagestyle{plain}
\begin{abstract}
IoT devices are increasingly the source of data for machine learning (ML) applications running on edge servers.  Data transmissions from devices to servers are often over local wireless networks whose bandwidth is not just limited but, more importantly, variable. Furthermore, in cyber-physical systems interacting with the physical environment, image offloading is also commonly subject to timing constraints. It is, therefore, important to develop an adaptive approach that maximizes the inference performance of ML applications under timing constraints and the resource constraints of IoT devices. In this paper, we use image classification as our target application and propose \textit{progressive neural compression ({\PNC})} as an efficient solution to this problem. Although neural compression has been used to compress images for different ML applications, existing solutions often produce fixed-size outputs that are unsuitable for timing-constrained offloading over variable bandwidth. To address this limitation, we train a multi-objective rateless autoencoder that optimizes for multiple compression rates via stochastic taildrop to create a compression solution that produces features ordered according to their importance to inference performance.  Features are then transmitted in that order based on available bandwidth, with classification ultimately performed using the (sub)set of features received by the deadline.  We demonstrate the benefits of {\PNC} over state-of-the-art neural compression approaches and traditional compression methods on a testbed comprising an IoT device and an edge server connected over a wireless network with varying bandwidth. 
\end{abstract}


\section{Introduction}
\label{sec:intro}
There is no denying the emergence of deep learning (DL) models broadens the scope of potential solutions for a variety of inference tasks in applications such as computer vision (CV)~\cite{krizhevsky2012imagenet, redmon2016you, he2016deep} and natural language processing (NLP)~\cite{devlin2018bert, graves2013speech}.  However, while much progress has been made in reducing the computational footprint of those models~\cite{sandler2018mobilenetv2, han2015deep}, they remain beyond the capabilities of most embedded systems, particularly low-end devices such as sensors.  As a result, bringing the capabilities of DL models to bear on tasks that arise in many Internet-of-Things (IoT) deployments has typically been realized through edge computing solutions. This setting is the focus of this paper that targets an image classification task performed in an edge server based on data acquired through a set of distributed sensors.


Although an advantage of edge computing is the relative proximity of compute resources with data sources (sensors), there is still a need for delivering the data from where they are captured to the edge servers.  When dealing with images, as we do in this paper, the sheer size of each data unit commonly calls for some form of compression due to the limited bandwidth of low-power networks such as Zigbee and LoRa.

Traditional image compression schemes are, however, designed with perceptual quality~\cite{hussain2018image} as their target, and, as we shall see, may perform poorly when the metric of interest is instead classification accuracy as in this paper.  This has led to much recent interest in compression schemes that are instead optimized for a given inference task. In particular, a number of solutions based on deep neural networks (DNNs) have been  proposed~\cite{eshratifar2019bottlenet, yao2020deep, matsubara2020head, shao2020bottlenet++, matsubara2022supervised} that explicitly incorporate inference accuracy in their design. 

These solutions, while effective, are limited by their inability to meet the timing constraint for image offloading under variable bandwidth. As real-time systems incorporate ML,  inference tasks often need to complete by a certain deadline, \eg the arrival of the next image or to allow actuation to be performed in time.  For example, consider a warehouse using image classification to automatically distribute packages to different lines. It is important for the image classification task to be completed before the arrival of the next package. The deadline for image classification in turn bounds the time available for image offloading. To meet offloading deadlines, it is essential for the system to \emph{adapt} to the bandwidth variations common to wireless networks~\cite{kulkarni20,mateo2019analysis}.  Under the varying bandwidth, the amount of data transmitted by the deadline varies.  The implication for any compression and inference scheme is, therefore, that it must be able to classify an image using only the amount of data actually received by the deadline.  More generally, the goal is to maximize classification accuracy given any amount of data received by the deadline.  Unfortunately, most prior approaches to neural compression generate a fixed amount of data, which must all be received for inference to be performed with sufficient accuracy. 

Towards addressing this limitation, we propose a framework called \textit{Progressive Neural Compression ({\PNC})}.
Our approach tackles the problem of progressive offloading by training a rateless compressive autoencoder (AE) that serves multiple objectives. Each objective focuses on optimizing the inference performance based on a different amount of data (partially) received, which corresponds to a different compression rate.
In our methodology, we leverage the ``taildrop'' technique~\cite{koike2020stochastic} in recent advances of neural dimensionality reduction to train the multi-objective AE that facilitates progressive image transmissions for classification. 
Specifically, classification can be performed at any time based only on the data received up to that point, with the receipt of more data translating into progressively more accurate classification decisions. This ``progressive'' property allows the system to adapt to variations in link bandwidth by continuously offloading data until reaching its offloading deadline.
This property is realized by training the AE to ensure that the latent features at its bottleneck layer are sorted in the order of their \emph{importance} for the classification task,
\eg earlier features have higher contributions towards accurate classification. Features are then transmitted in the order of their importance till the deadline. Classification can be performed on the edge server based on the subset of received features, with the classification accuracy improving as more features are received. The progressive nature of {\PNC} allows it to handle different deadlines and varying bandwidth effectively. 


The main contributions of this paper are as follows:
\begin{itemize}
    \item We propose {\PNC}\footnote{{The source code and testbed setup instructions for {\PNC} are available at: \url{https://github.com/rickywrq/Progressive-Neural-Compression}.}}, an efficient progressive neural compression framework tailored for IoT devices that need to offload images to edge servers for classification within a deadline despite varying network bandwidth. 
    
    \item We designed and implemented an end-to-end image classification system integrating time-bounded progressive image offloading and edge-based classification on an edge computing testbed.

    \item We demonstrated empirically that {\PNC} outperformed both traditional and neural compression solutions in classification accuracy under timing constraints and external wireless interference while realizing a much smaller computational footprint than the latter.  
\end{itemize}


\section{Background and Motivation}
\label{sec:background}



IoT devices are usually connected to edge servers by low-power wireless sensor networks (e.g., IEEE 802.15.4 or LoRa) with limited bandwidth. Before offloading an image from the device to the server, the image needs to be compressed while still allowing the edge server to classify it accurately. Moreover, these networks typically experience significant fluctuations in available bandwidth due to environmental factors and interference in the wireless channel~\cite{10.1145/1062689.1062741, 10.1145/1689239.1689246, 10.1145/958491.958493}. Furthermore, image classification in such settings is often subject to timing constraints.  For example, a quality control camera needs to catch a defective item before the arrival of the next item, or a face recognition camera needs to react quickly to prevent unauthorized entry.


The combination of variable and limited bandwidth and time constraints has important implications for image offloading.  It impacts the amount of data that can be transmitted before classification has to be made. As a result, any edge-based image classification system must be able to adapt to fluctuations in the available bandwidth. One option is to rely on a feedback mechanism that detects bandwidth variations and responds by adjusting compression ratios accordingly. The main disadvantage is the inherent latency associated with having to detect variations before being able to react to them. As a result, we focus on ``open-loop'' solutions that optimize classification performance given the amount of data transmissions the network allows by the deadline. 

We identify the following requirements for an image offloading framework for such image classification:

\begin{itemize}[nosep]
    \item \textbf{Classification accuracy:} In contrast to traditional image compression approaches optimized for image reconstruction, an offloading framework designed for image classification should target classification accuracy as the primary objective of interest to the application. 
    
    
    \item \textbf{Encoding efficiency:} Given the stringent resource constraints of IoT devices, the encoding process of the compression approach should incur minimal computation overhead on the devices.
  
    \item \textbf{Adaptation to different deadlines and varying bandwidth:} 
    Classification accuracy should improve the more data are received by the deadline while achieving graceful performance degradation in the face of decreasing bandwidth.
    

\end{itemize}

\section{Related Work}

Image compression aims at minimizing the size of image representation by removing redundant and irrelevant information~\cite{hussain2018image}. In this section, we review related works on image compression, with a focus on the requirements introduced in Section~\ref{sec:background}.



\subsection{Traditional Image Compression} 
Traditional compression techniques aim to minimize the amount of data needed to reconstruct an image. For example, JPEG~\cite{wallace1992jpeg, pennebaker1992jpeg}, one of the most widely used image compression standards, works by performing discrete cosine transform (DCT) on the original image and then entropy coding the quantized DCT coefficients. WebP~\cite{webpurl} is a more recent image compression standard commonly used on the web. Both the basic JPEG standard and WebP are \textit{non-progressive} in that they require the reception of an entire compressed image to reconstruct it effectively. In contrast, to handle unpredictable bandwidth, JPEG also provides a \emph{progressive} mode based on spectral selection and successive approximation in the entropy coding procedure. It allows images to be reconstructed based on an arbitrary amount of encoded data, with more data resulting in more accurate reconstruction.


\subsection{Neural Compression for Image Reconstruction} 
Recent works in image compression explored deep learning approaches using AEs to learn latent representations of images. 
Early efforts on such neural compression techniques ~\cite{theis2017lossy, balle2018variational, minnen2018joint} are non-progressive because their neural network architecture cannot handle partially received data. Recent efforts are to support \emph{progressive} encoding in neural compression. Lu \etal~\cite{lu2021progressive} propose nested quantization with multiple scaling levels that refines all latents progressively. Toderici \etal~\cite{toderici2015variable, toderici2017full} exploit recurrent neural networks (RNN) to support progressive image compression, where a multi-iteration compression architecture is proposed to iteratively compress the residual signals between the input and reconstructed image patch. While neural compression has shown advantages over traditional compression approaches in compression ratio, a disadvantage of existing deep models for image compression is their complexity and computation cost, often prohibitive for resource-constrained IoT devices. It is also important to note that the aforementioned neural compression approaches are designed to optimize image reconstruction quality instead of application-specific inference performance such as image classification accuracy.

\subsection{Neural Compression for Inference} 
Recent advances in neural compression target optimizing inference performance. 
In the context of edge computing, several works propose approaches where the device and the edge server work collaboratively to complete the inference task~\cite{kang2017neurosurgeon, eshratifar2019jointdnn}. In this paradigm, a DNN model for inference is split into two separate parts, a head model and a tail model, which are deployed at the device and the edge server, respectively. 
To compress the intermediate features transmitted from the device to the edge server, BottleNet and its extensions~\cite{eshratifar2019bottlenet, shao2020bottlenet++} inject a bottleneck AE at the split point of the original model. 
To improve encoding efficiency, Yao \etal~\cite{yao2020deep} propose an asymmetric AE structure to lower the encoding overhead at the local device.  To further reduce redundant computation, Matsubara \etal~\cite{matsubara2019distilled, matsubara2020head} apply knowledge distillation to train a lightweight head model with an embedded bottleneck that compresses the input image while performing part of the inference.  In subsequent work~\cite{matsubara2022supervised}, this head distillation framework was extended to achieve an explicit learnable trade-off between compression ratio and reconstruction distortion by combining neural image compression techniques~\cite{balle2018variational, alemi2016deep, singh2020end}.  However, despite the significant advances in neural compression for inference, none of these solutions support progressive compression, which is essential in adapting to variable bandwidth, as is common in wireless sensor networks.


A recent neural compression approach, Starfish (Hu \etal)~\cite{10.1145/3384419.3430769}, accounts for data loss during transmission by introducing random dropouts at the bottleneck of their proposed AE to foster resiliency against such losses.  However, as Starfish does not differentiate or prioritize features in terms of their contributions to inference accuracy, it is not designed to optimize classification performance under different deadlines and available bandwidth.

Our proposed solution, {\PNC}, is specifically designed to optimize image classification accuracy in the context of edge offloading under timing constraints. In contrast to existing solutions, {\PNC} can adapt to variable bandwidth through progressive encoding. It lets the edge server perform effective image classification based on the subset of encoded features that have been received by the offloading deadline. In addition, to minimize computational load on IoT devices, {\PNC} employs a lightweight rateless AE and does not require inference to be partially executed on the devices.

\begin{figure*}[t]
    \centering
    \includegraphics[width=0.9\linewidth]{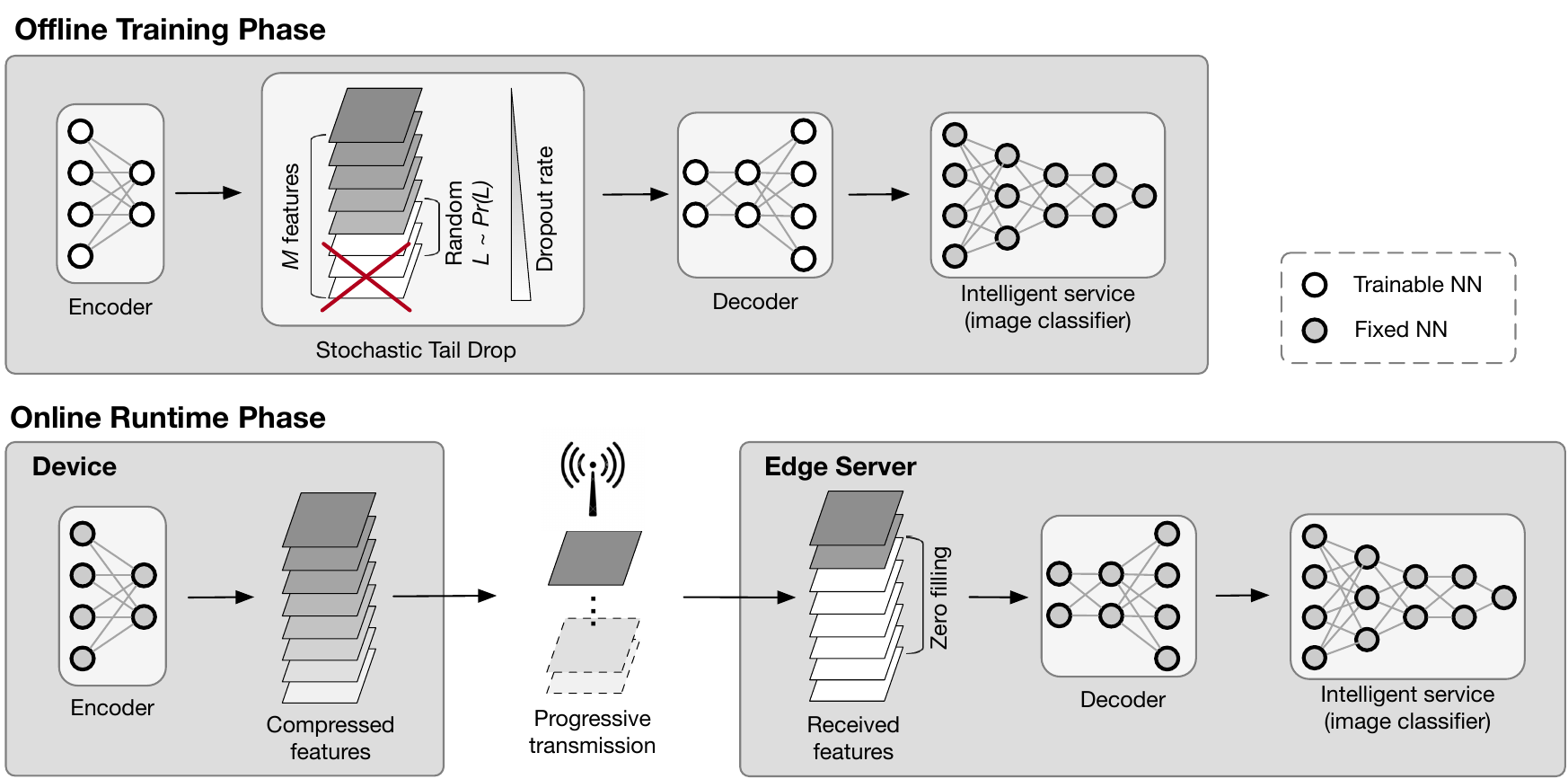}
    \caption{System overview of our {\PNC} offloading framework. }
    \label{fig:system_overview}
\end{figure*}

\section{Problem Formulation}
\label{sec:problem_formulation}
We consider an edge image classification task where images are captured in a local device and must be transmitted to an edge server for classification purposes. Because of bandwidth limitations, images are encoded (compressed) before transmission and subsequently decoded at the edge server before classification. Our goal is to achieve the highest classification accuracy possible at the edge server, given the bandwidth constraints of the network and the image offloading deadline. We focus on the timing constraint for image offloading subject to variable bandwidth. 

Assume a sequence of images, $\mathbf{x}_i$, $i=1,2,3,\dots$, captured by the local device at fixed time intervals of period $T$. Assume the first image is captured at $t_1$, then $\mathbf{x}_i$ will be captured at $t_i = t_1 + (i-1)T$. In this work, we consider a common scenario where the image offloading must be completed before the next image is encoded, i.e., the deadline for offloading an image $\mathbf{x}_i$ is the arrival of the encoded features of the next image $\mathbf{x}_{i+1}$.

Specifically, we split the process into three steps: On-device encoding and network offloading, and subsequently edge decoding and classification. 

\textbf{On-device Encoding.}
Upon the arrival of an image $\mathbf{x}_i$, the device encodes and sends the data to a remote server for image classification. Given an encoder, $f_\theta$, the encoded data for each image is in the form of a byte array that can be expressed as, $\mathbf{z}_i = f_\theta(\mathbf{x}_i)$.
The encoding latency for $\mathbf{x}_i$ 
is denoted as $t^f_{i}$.

\textbf{Network Offloading.}
We assume that the transmission of image $\mathbf{x}_i$ can continue until the next image $\mathbf{x}_{i+1}$ is encoded and ready for transmission, so that its deadline is $t_d = t_i+T+t^f_{i+1}$. As the network bandwidth, $b(t)$, varies with time,
the encoded data size $S_i$ that can be transmitted for image $\mathbf{x}_i$ by its deadline is given by
$S_i = \int_{t_i+t^f_{i}}^{t_i + T+t^f_{i+1}} b(t)dt$.

Denote $\mathtt{truncate}(\mathbf{z}_i,S_i)$ as the data when truncating the end of byte array $\mathbf{z}_i$ to size $S_i$, such that $\mathtt{size}\left(\mathtt{truncate}(\mathbf{z}_i,S_i)\right)\leq S_i$ (note that when $\mathtt{size}(\mathbf{z}_i)\leq S_i$, $\mathtt{truncate}(\mathbf{z}_i,S_i) = \mathbf{z}_i$). The actual data, $\tilde{\mathbf{z}_i}$, that the client can send within the deadline is,
\begin{equation}
    \tilde{\mathbf{z}_i}=  \left\{ 
  \begin{array}{ l l }
    \mathbf{z}_i,                            & \text{if } \mathtt{size}({\mathbf{z}_i}) \leq S_i, \\
    \mathtt{truncate}\left(\mathbf{z}_i,S_i\right),     & \textrm{otherwise}.
  \end{array}
\right.
\end{equation}

\textbf{Decoding and Classification.}
When the edge server receives the (potentially truncated) encoded data, $\tilde{\mathbf{z}_i}$, the decoder, $g_\phi$, will decode the data, $\mathbf{y}_i=g_\phi(\tilde{\mathbf{z}_i})$,
Assume that $h$ is some image classifier and $a_i$ is the output of the image classifier, \ie the probability distributions among the possible classes with input $\mathbf{y}_i$, $a_i=h(\mathbf{y}_i)=h\left(g_\phi(\tilde{\mathbf{z}_i})\right)$.
We can evaluate the accuracy of the prediction with the commonly used performance metric \emph{top-$n$ accuracy}, $R$, of the image classification task on image $\mathbf{x}_i$ with ground-truth class of $gt_i$ as,

\begin{equation}
\hspace{-0.25cm}
R(a_i, gt_i)
=  \left\{ 
      \begin{array}{ l l }
        1, & \text{if $gt_i\in$\{$n$ most likely predictions\},} \\
        0, & \text{if $gt_i \notin$\{$n$ most likely predictions\}.}\\
      \end{array}
  \right.
\label{eqn:problem_formulation:top_n_acc}
\end{equation}

Our goal is to design a flexible configuration for edge offloading and optimize the learnable parameters of the image encoder and decoder, $\theta$ and $\phi$, that maximize the accuracy of the image classification task, under variable bandwidth $b(t)$ and the constraint of the deadline imposed by $T_d$.

In summary, our goal is to design an offloading solution that can achieve high accuracy with various combinations of time-varying bandwidth and deadline, \ie an unknown amount of transmitted data in an adaptive fashion. The encoder should capture important features for classification and help the client progressively transmit the most important features first, followed by other, less important features. The decoder should handle incomplete data arriving at the edge server and decode an input for the image classifier to achieve the best classification accuracy given the amount of data received.

\section{Design of {\PNC}}

This section introduces the design of our {\PNC} framework for progressive neural compression and adaptive offloading. We target a typical real-time edge computing architecture comprising embedded IoT devices and an edge server connected by a wireless sensor network, where image offloading is regulated by a deadline constraint. The key idea of our framework is to formulate the problem of progressive offloading as training a rateless compressive AE that serves multiple objectives, where each objective corresponds to optimizing the inference performance using a certain amount of data partially received during transmission.
We exploit recent advances in dimensionality reduction for multi-rate image compression and extend such a concept to support progressive offloading of inputs for edge-assisted image classification, thereby allowing the system to adapt to varying network bandwidth under timing constraints.



\subsection{System Overview}
\label{sec:pnc_overview}


Fig.~\ref{fig:system_overview} illustrates the offline training process and the end-to-end image classification pipeline featured by the {\PNC} framework. During the offline training phase, the {\PNC} trains a rateless compressive AE by dropping a random number of encoded features (by filling these features with zeros) at the tail of its bottleneck during each training iteration to emulate the reception of incomplete image data (Section \ref{sec:pnc} and \ref{sec:system_design:two_stage}). When the {\PNC} is later deployed at runtime, the device encodes the image into a fixed number of features and transmits to the edge server as many features as possible within the time limit. With the features that failed to be transmitted in time replaced with zeros, the edge server decodes the received data and predicts the class of the input image (Section \ref{sec:offloading}).

\subsection{Multi-objective Neural Compression}
\label{sec:pnc}
\subsubsection{AE-based Fixed-rate Compression}
A vanilla compressive AE consists of an encoder and decoder. The encoder, denoted as $\mathbf{z} = f_{\theta}(\mathbf{x})$, maps the input image $\mathbf{x}\in\mathbb{R}^{H\times W\times C}$ (H, W, and C define the size and number of channels of the input. For RGB images, $C=3$.) to a lower-dimensional latent representation $\mathbf{z}\in \mathbb{R}^{h\times w \times M}$ ($h$ and $w$ define the size of the latent feature map for each channel, $M$ is the number of latent channels). $\theta$ is the parameters of the encoder. The decoder, denoted as $\hat{\mathbf{x}} = g_{\phi}(\mathbf{z})$, aims to reconstruct the original image $\mathbf{x}$ by mapping the latent representations back to the data space. $\phi$ is the parameters of the decoder. The aim is to find $\theta,\phi$ that jointly minimize a loss:
\begin{equation}\label{eq:loss}
\theta, \phi = \operatorname*{arg\,min}_{\theta,\phi}\mathcal{L}\left(\theta, \phi\right)
\end{equation}
with $\mathcal{L}(\theta, \phi) = \mathbb{E}_{\textbf{x}\sim \mathcal{X}}\|\mathbf{x} - g_{\phi}\left(f_{\theta}(\mathbf{x})\right)\|^2$ optimizing the AE to reconstruct the input images and minimize the distortion~\cite{theis2017lossy, balle2018variational, minnen2018joint}.
In compressive offloading~\cite{yao2020deep,matsubara2020head}, however, the AE is often trained to optimize inference performance (e.g., classification accuracy). Here the loss $\mathcal{L}$ can be generalized to inference losses (e.g., cross entropy) to meet the needs of applications.
The natural property of dimensionality reduction of AE enables its potential of being used in compressive offloading for edge computing~\cite{matsubara2022supervised,matsubara2020head,yao2020deep} where $\mathbf{z}$ is transmitted as compressed data to improve efficiency. 

\subsubsection{Multi-objective Optimization for Rateless Compression}
The problem with using a standard AE for compressive offloading is that the compression rate remains fixed due to the fixed-sized bottleneck $\mathbf{z}$. In wireless sensor networks with varying bandwidths and offloading deadlines, the bottleneck features may not be fully received. This leads to significant distortion and performance decline for the target deep learning service, as the bottleneck features are always complete during training.

To reduce the distortion and optimize  classification performance under the situation where only a subset of the bottleneck features are successfully received by the edge, \ie the first $K$ features ($K \leq M$), we can incorporate the objective w.r.t. reconstructing data using the top $K$ channels,
\begin{equation}\label{eq:oneloss}
    \mathcal{L}(\theta, \phi, K) = l\left(\mathbf{x}, g_{\phi}\left(\text{Concat}\left[\mathbf{z}_{[1:K]}; \mathbf{0}\right]\right)\right)
\end{equation}
where $l$ represents the choice of objective function to optimize during the two-stage offline training process, which we will discuss in Section~\ref{sec:system_design:two_stage}, $\text{Concat}([;])$ denotes tensor concatenation, and $\mathbf{z}_{[1:K]}$ denotes the first $K$ latent feature channels of $\mathbf{z}$.
To consider all possible $K$ values under different bandwidth, we can train a AE $(f_\theta, g_\phi)$ by optimizing the following \textit{multi-objective} problem:
\begin{equation}\label{eq:moloss}
\min\limits_{\theta,\phi} \mathop{\mathbb{E}}\limits_{\textbf{x}\sim \mathcal{X}} [{\mathcal{L}}(\theta,\phi;1),{\mathcal{L}}(\theta,\phi;2),...,{\mathcal{L}}(\theta,\phi;M)]
\end{equation}
However, there are no close-form solutions to Eq.~(\ref{eq:moloss}).
Since each objective is not independent of each other (i.e., sharing the model parameters $\theta$ and $\phi$), the rateless AE must account for the trade-off among various bottleneck dimensionality to reach the \textit{Pareto frontier} solutions as close as possible.

Recent work~\cite{koike2020stochastic} proposed the \textit{stochastic taildrop} regularization as an iterative approximation to the solutions of Eq.~(\ref{eq:moloss}) for flexible neural dimensionality reduction. As illustrated in the ``Offline Training Phase'' diagram in Fig.~\ref{fig:system_overview}, in each training step, we randomly drop the last $L$ latent channels ($L = M - K$) of the bottleneck by replacing them with $0$ during training. 
Intuitively, this is equivalent to alternatively optimizing each of the objectives with different $K$ at different iterations. On top of \textit{stochastic taildrop}, we added additional gradient control over the iterations for more stable training.
The iterative joint optimization is shown in Algorithm~\ref{alg:iter}. 
At each training iteration (the innermost for loop), we randomly draw the length of dropped tail $L$ from a distribution~\cite{koike2020stochastic} (e.g., uniform) and calculate the gradient $\mathbf{g}_{\theta}^{(m)},\mathbf{g}_{\phi}^{(m)}$
following the objective $\mathcal{L}$.
After every $M$ iterations, the training process will calculate the average gradient and update the learnable parameters, $\theta$ and $\phi$.

\RestyleAlgo{ruled}
\SetKwComment{Comment}{/* }{ */}

\begin{algorithm}[t!]
\fontsize{8.5pt}{10pt}\selectfont
\caption{\small Iterative optimization of Eq. (\ref{eq:moloss})  with stochastic taildrop.}\label{alg:two}
\KwData{Number of epochs $N_e$, number of batches $N_b$, mini-batch image set $\mathcal{X}_{\textit{train}}=\left\{\mathbf{X}_1, \mathbf{X}_2, ..., \mathbf{X}_{N_b}\right\}$, predefined distribution $\mathit{Pr}(L)$, bottleneck width of the AE $M$}
\KwResult{Learnable parameters of the encoder, $\theta$, and the decoder, $\phi$.}
\For{$1 \leq n_e \leq N_e$ }{
  \For{$1 \leq n_b \leq N_b$}{
    \For{$1 \leq m \leq M$}{
    $L \sim \mathit{Pr}(L)$ \Comment*[r]{\scriptsize{Length of dropped tail.}}
    
    $K \gets M - L$; 
    
    Gradient $\mathbf{g}_{\theta}^{(m)},\mathbf{g}_{\phi}^{(m)} \gets \nabla_{\theta,\phi} \mathop{\mathbb{E}}\limits_{\textbf{x}\sim \mathbf{X}_{n_b}} {\mathcal{L}}(\theta,\phi;K)$;
    }
    
    $\bar{\mathbf{g}}_{\theta}, \bar{\mathbf{g}}_{\phi} \gets \frac{1}{M} \sum_{m=1}^{M} \mathbf{g}_{\theta}^{(m)}$, $\frac{1}{M} \sum_{m=1}^{M} \mathbf{g}_{\phi}^{(m)}$; \\
    
    Update $\theta$, $\phi$ using gradient $\bar{\mathbf{g}}_{\theta}$, $\bar{\mathbf{g}}_{\phi}$ on batch $\mathbf{X}_{n_b}$.
  }
}
\label{alg:iter}
\end{algorithm}

\subsection{Two-stage Offline Training}
\label{sec:system_design:two_stage}
In this section, we introduce how to optimize our rateless AE for downstream inference tasks (e.g., image classification) via a two-stage training method.

\subsubsection{Unsupervised Pretraining for Image Reconstruction}
We first pre-train the rateless AE by optimizing Eq. (\ref{eq:moloss}) via Algorithm \ref{alg:iter} on reconstructing input images $\mathbf{x} \sim \mathcal{X}$, in which 
\begin{equation}
    \mathcal{L}(\theta, \phi, K) = \left|\left|\mathbf{x} - g_{\phi}\Big(\text{Concat}\left[f_{\theta}(\mathbf{x})_{[1:K]}; \mathbf{0}\right]\Big)\right|\right|^2
\end{equation}
The unsupervised pretraining helps prepare the AE parameters $\theta$, $\phi$ to a generalizable state that adapts to any potential target machine learning (ML) service.

\subsubsection{Knowledge Distillation for Inference}\label{sec:knowledge_distill}
The parameters $\theta$, $\phi$ trained by the above unsupervised stage are not optimized for inference performance (e.g., classification accuracy).  
In order to achieve a better performance-efficiency tradeoff for inference, the pretrained rateless AE needs to further distill knowledge from the target inference task. Knowledge distillation (KD) was first proposed in~\cite{hinton2015distilling} and widely used by ~\cite{yao2020deep,matsubara2019distilled,matsubara2022supervised} and showed improved inference performance than training from scratch. 
In KD, a (usually smaller) student model is trained to produce similar inference output as the (usually much larger) target teacher model. 
In this paper, we consider a deep image classifier to instantiate $h(\mathbf{x})$ and extend the scope of stochastic taildrop from image compression and reconstruction to image classification. Assume the deep model of the target inference task is $h(\mathbf{x})$. We use KD to fine-tune the rateless AE to reconstruct input data that optimize the inference performance with the new objective function specifically for the classification task,
\begin{equation}
    \mathcal{L}(\theta, \phi, K) = \mathrm{CE} \left( h(\mathbf{x}), h\Big( g_{\phi}\Big(\text{Concat}\left[f_{\theta}(\mathbf{x})_{[1:K]}; \mathbf{0}\right]\Big)\Big) \right)
\end{equation}
Here $\mathrm{CE}(\cdot, \cdot)$ denotes cross-entropy loss as used in \cite{hinton2015distilling}.
During the KD training, the target model $h(\mathbf{x})$ is treated as a black box with the parameters frozen, and only the AE parameters, $\theta$ and $\phi$, are updated. 

\subsection{Progressive Compression for Adaptive Offloading}
\label{sec:offloading}
With the stochastic taildrop regularization and iterative optimization (Algorithm~\ref{alg:iter}) during training, 
the $p^\text{th}$ feature $\mathbf{z}_{[p]}$ is preserved and used for reconstruction more often than any succeeding feature
$\mathbf{z}_{[q]} (1 \leq p < q \leq M)$. 
Intuitively, the AE will be more robust to the absence of $\mathbf{z}_{[q]}$, so features ${\mathbf{z}}_{[1:M]}$ are given decreasing importance to the target inference task performance.

As shown in the diagram of ``Online Runtime Phase'' in Fig.~\ref{fig:system_overview}, we design an end-to-end image classification system to support progressive neural compression and image offloading over wireless
sensor networks. After encoding original images into a set of compressed features, we progressively transmit the features from the most important $\mathbf{z}_{[1]}$ to the least important $\mathbf{z}_{[M]}$ in descending order to the edge server for further reconstruction and inference. Given limited current bandwidth and a time constraint, the network may be able to transmit only a subset of the features (i.e., top $K'$) to the edge server before the deadline. Then the received most important $K'$ features will be used for data reconstruction and inference (remaining features filled with 0). These received features are exactly the optimal set of features that maximizes the inference performance of the image under the constraint of sending only $K'$ features. In this manner, the trained rateless AE achieves progressively better inference performance with more features offloaded. This progressive neural compression framework enables our compressive AE to adapt flexibly to variable network bandwidth while achieving better average inference performance.

\subsection{Quantization and Huffman Coding}\label{sec:quant_huffman}
To further reduce the data size of intermediate features $\mathbf{z}$ that need to be transmitted to the edge server, we apply additional optimizations, including model and feature quantization and Huffman coding~\cite{huffman1952method} before data transmission. 

\textbf{Model Quantization.} Model quantization~\cite{polino2018model, zhou2018adaptive} converts floating point operations and data in ML models to quantized ones, usually in the form of 8-bit integers. We use the automated TensorFlow Lite (TFLite) converter, an official tool by TensorFlow, for such a purpose. The conversion speeds up the encoding on resource-constrained IoT devices and helps reduce the output data size of the ML model. 

\textbf{Feature Quantization.} We empirically find that even after model quantization to 8-bit integers (0-255), there is still abundant room for more aggressive quantization. After encoding, we further quantize the data to 64 values, represented by 6 bits, spaced evenly between 0 and 255. We select the number 64 because it has a relatively small impact on the classification accuracy while directly reducing the data size by $25\%$ if we use 6-bit binary coding.

\textbf{Feature-wise Huffman Coding.} We find that the data distributions within each feature are different. For the best compression result, we train $M$ different Huffman tables for features at different positions. During the online runtime phase, each quantized feature will be further compressed individually with the corresponding Huffman table.

\subsection{Progressive Property of PNC}

{
In this section, we provide insights into how the design of {\PNC} facilitates its progressiveness and adaptability to bandwidth variations with little impact on its classification performance. To better demonstrate the effect of taildrop on multi-objective rateless compression, we remove tail-drop regularization from {\PNC} and vary the number of latent channels, $M$, in the model to create multiple \textit{fixed-rate AE}s. The bottleneck sizes of these fixed-rate AEs are $32\times32\times K$, where $K\in \{2,4,6,8,10\}$ represents the latent channel numbers and can effectively translate to various encoding rates.  All fixed-rate AEs are trained and optimized following the same strategy as {\PNC}, as discussed in Section~\ref{sec:system_design:two_stage} and Section~\ref{sec:offloading}. As a result, each latent channel will be of size 1KB after model quantization and become 0.75KB after feature quantization. Huffman coding will further improve coding efficiency and reduce the average channel size. We utilize the ImageNet validation set with 35,000 images for training, 5,000 for validation, and 10,000 for the following simulation evaluation.

\begin{figure}[!t]
    \centering
    \includegraphics[width=0.85\linewidth]{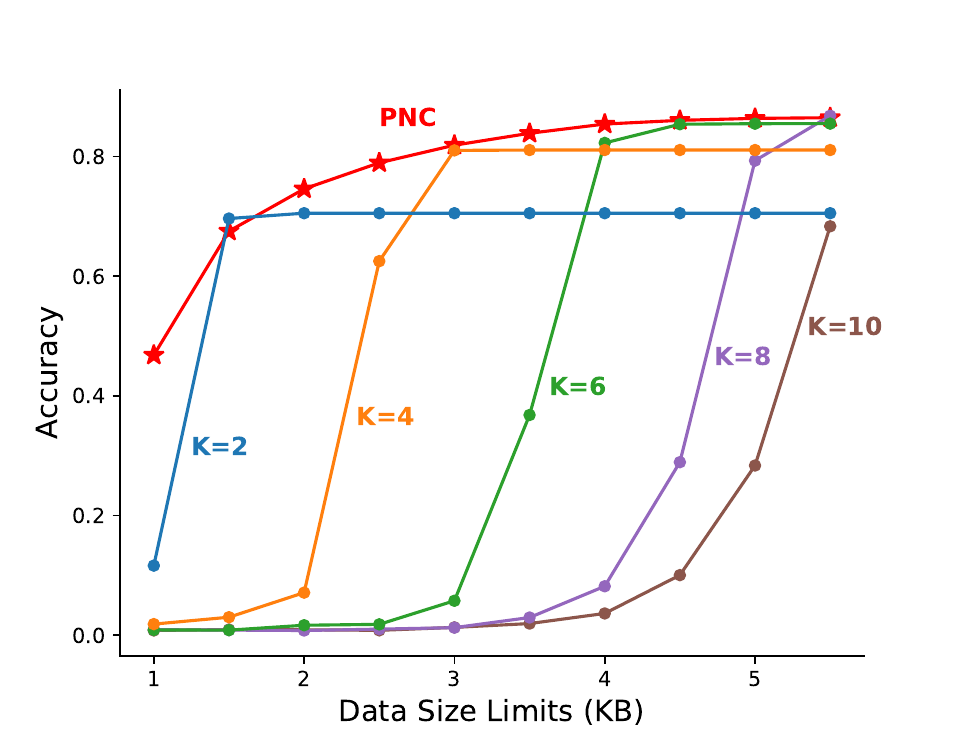}
    \caption{
    Classification accuracy vs data size limits of {\PNC} and various fixed rate AEs. {\PNC} is in red. For fixed-rate AEs, a label with the number of latent channels, $K$, is next to each curve with the corresponding color.
    }
    \label{fig:simu:acc_dsize_fixed_rate_AEs}
\end{figure}

Fig.~\ref{fig:simu:acc_dsize_fixed_rate_AEs} shows, as a function of data size limitation, the classification accuracy of {\PNC} and those fixed-rate AEs. First, we note that {\PNC} consistently delivers classification accuracy close to the \textit{best} performance of each fixed-rate AE (even though the fixed-rate AE is optimized for the target size of latent features). The slight drop in performance in comparison to the best fixed-rate AE at each data size is attributed to the constraints imposed by stochastic taildrop during training (\eg the accuracy for $K=2$ is $69.62\%$, just slightly higher than {\PNC}'s $67.51\%$ at 1.5KB. The accuracy for $K=8$ is $86.79\%$, barely above {\PNC}'s $86.46\%$ at 5.5KB.). Overall, the multi-objective design enabled by stochastic taildrop during training allows {\PNC} to balance the inference performance under multiple compression rates during inference. PNC is therefore able to achieve progressive encoding at a negligible cost in classification accuracy across a wide range of data sizes. In other words, the accuracy of {\PNC} gracefully degrades as the number of offloaded features decreases.

In contrast, the performance of fixed-rate AEs decreases rapidly as the data size limit drops substantially from their respective design targets, indicating a failure of the decoder to receive the complete set of features.  Without stochastic taildrop, each feature in a fixed-rate AE will be equally important and indispensable for decoding and classification. This becomes a serious problem for DNNs-based image encoders with fixed rates when the amount of data sent to the server is unknown or varies. Similarly, selecting the best model that guarantees the transmission of all encoded data within a time limit becomes challenging. Therefore, a significant advantage of {\PNC} lies in its progressive encoding that enables the system to adapt to unpredictable bandwidth and different offloading deadlines while maintaining classification accuracy.
}

\subsection{Image Offloading Architecture}
\begin{figure}[t]
    \centering
    \includegraphics[width=0.9\linewidth]{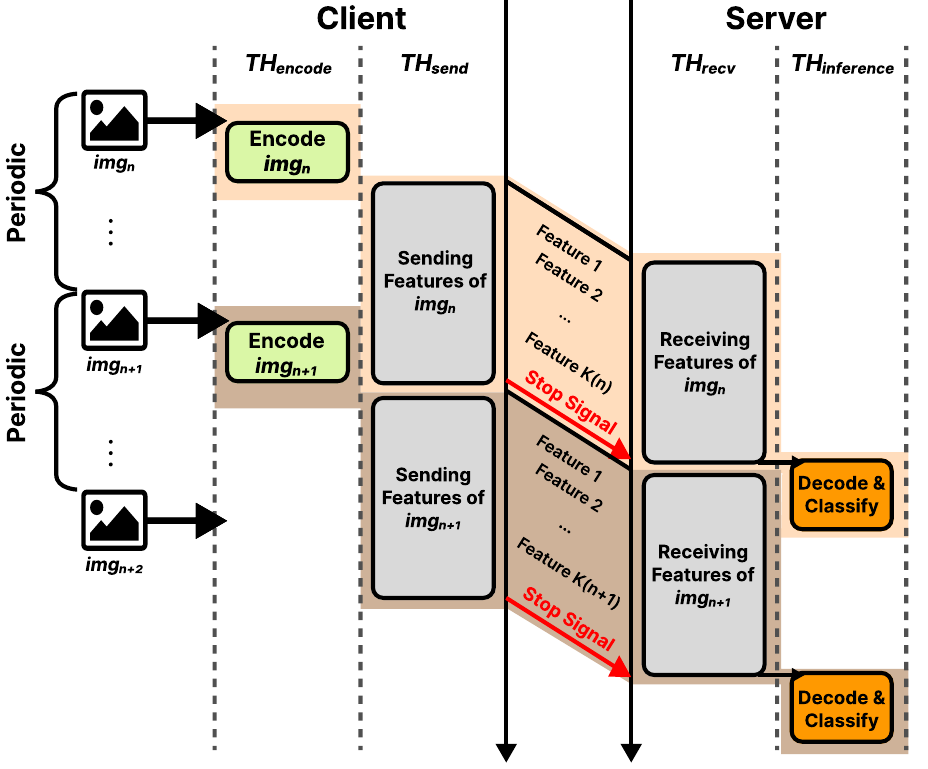}
    \caption{The image offloading architecture in our end-to-end image classification system. 
    }
    \label{fig:offloading_schedule}
\end{figure}

We design a distributed system architecture (Fig.~\ref{fig:offloading_schedule}) for end-to-end image classification comprising multiple stages: image encoding and offloading on the IoT device (client), followed by decoding and classification on the edge server. Recall our design targets a common application scenario where the image classification task is triggered by images captured periodically. We focus on the latency of the image offloading stage subject to varying bandwidth. In our design, the image offloading stage has a time constraint that the transmission of features belonging to an image must end before features of the next image are generated.

To improve resource utilization, our system employs multiple threads to pipeline the stages of the end-to-end processing. Specifically, image encoding is released in response to the arrival of a new image. Thread $\mathit{TH}_\textit{encode}$ first fetches and encodes the image into compressed features. Thread $\mathit{TH}_\textit{send}$ then sends the features to the edge server over the network.
{To provide fine-grained control, $\mathit{TH}_\textit{send}$ divides the features of an image into small, 64-byte data blocks and sends them sequentially. Before sending each block, $\mathit{TH}_\textit{send}$ will check whether the encoded data of the next image are ready.}
If the transmission of $\mathit{img}_n$ features is not finished before the first feature of $\mathit{img}_{n+1}$ is encoded, $\mathit{TH}_\textit{send}$ terminates the transmission, {appends a stop signal to the data stream indicating the termination of transmission for $\mathit{img}_n$,} and proceeds to transmit $\mathit{img}_{n+1}$. 

{\PNC} uses Huffman encoding as the final optimization step after an image is compressed into features. To reduce the latency overhead caused by the Huffman encoding, $\mathit{TH}_\textit{send}$ is invoked to offload the first feature once it is encoded, while the remaining features are encoded with the Huffman table in parallel using a third thread (not shown in the figure).

The edge server runs two threads: Thread $\mathit{TH}_\textit{recv}$ receives the bytes from the network interface until it receives a complete image or the stop signal. Then thread $\mathit{TH}_\textit{inference}$ decodes and classifies $\mathit{img}_n$ using the received data. Our architecture is implemented in Linux-based OSs on both the device and the server, although, in principle, it may be implemented in other OSs supporting multiple threads.

\section{Implementation}
We implement an end-to-end image classification system comprising an IoT device and an edge server. The system employs {\PNC} for image offloading over a low-power wireless sensor network. 


\subsection{Hardware Platform}
\label{sec:implementation:hw}
We use a Raspberry Pi 4 Model B with Quad core Cortex-A72 (ARM v8) 64-bit SoC @ 1.5GHz, 8GB RAM, and Raspberry Pi OS 10 (Kernel version: 5.10.63) as the embedded IoT device and a desktop with Intel(R) Core(TM) i7-10700K CPU @ 3.80GHz, Nvidia GeForce RTX 3090 GPU, 64GB RAM and Ubuntu 20.04.1 (Kernel version: 5.15.0-69-generic) as the edge server. The device and the server communicate with each other using IEEE 802.15.4 radios (nRF52840 development kits ``nRF52840 DK'' produced by Nordic Semiconductor). The kit offers a serial port interface that allows the client and the server to transmit and receive data streams with the IEEE 802.15.4 radio at the application level. The MAC layer of the radio uses re-transmissions to overcome transmission failures. As a result, the available network bandwidth drops in the presence of external inference and packet losses.

\begin{figure*}[!t]
    \centering
    \includegraphics[width=0.95\linewidth]{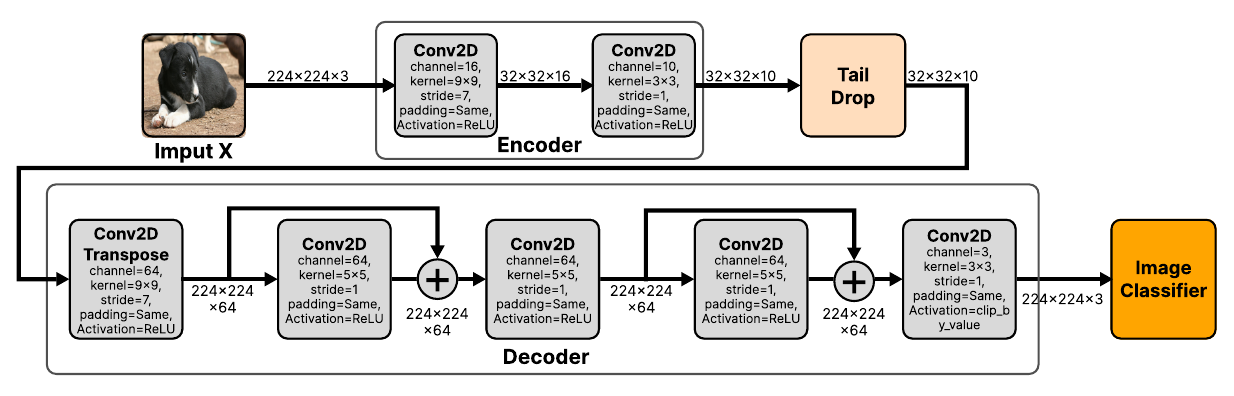}
    \caption{The AE architecture used in PNC. The encoder contains two 2D convolution layers. The decoder contains five layers. The input/output size of each layer is labeled between layers.}
    \label{fig:model_architecture}
\end{figure*}

\subsection{Design Choice of Autoencoder}
We implement the AE in {\PNC} with TensorFlow\footnote{\url{https://www.tensorflow.org}}. 
The details of our compressive AE are shown in Fig.~\ref{fig:model_architecture}. Inspired by Yao \etal~\cite{yao2020deep}, we adopt an asymmetric design of the AE to reduce the encoding overhead at the device. The encoder and decoder contain two and five layers, respectively. The encoder compresses the input image into 10 features of size $32\times 32$. We make the encoder lightweight, \ie with only two 2D-Convolution (Conv2D) layers to match the computational capacity of the embedded device. We use clip-by-value as the activation function in the last layer to limit the AE output within a reasonable range. Empirically, we find that {\PNC} trained with clip-by-value achieves higher classification performance than using the common sigmoid or tanh activation functions.
For tail length selection,  we use the uniform distribution following~\cite{koike2020stochastic}, \ie $L \sim \texttt{Uniform} (0,M-1)$, and zero out the last $L$ channels during training. 

We train the AE in {\PNC} following the steps in Section~\ref{sec:system_design:two_stage}. We use a learning rate of 0.001 for the first stage and 0.005 for the second. We set the batch size to 4 and the number of epochs to 70, and use the Adam optimizer during both training stages.

We use ImageNet\footnote{\url{https://www.image-net.org/}.} for model training and system evaluation. We use the off-the-shelf EfficientNet-B0~\cite{48187} pre-trained on the ImageNet training set as our image classifier. Since the training set is already used to optimize the image classifier, we extract additional training data from the ImageNet validation set to train {\PNC}'s AE. We assigned 35,000 images for training, 5,000 for validation, and 2,000 for testbed experiments. We also resize all the images to $224\times 224\times 3$ to match the input size of the classifier.

\section{Evaluation}

In this section, we compare the encoding efficiency and classification performance of {\PNC} against a set of traditional and state-of-the-art baselines under different timing constraints (deadlines) and network conditions.

\subsection{Baselines for Comparison}
\label{sec:evaluation:baselines}
We compare {\PNC} against a wide range of baselines, including traditional and neural compression approaches, as well as non-progressive and progressive encoding schemes of different levels of encoding complexity. 
\begin{itemize}

    
    \item \textbf{WebP:} WebP is a state-of-the-art, non-progressive, traditional encoder. In addition to the quality factor, $q$, that controls the reconstruction quality of the image, WebP has another parameter, \textit{method}, which is an integer ranging from 0 to 6, indicating the quality-speed trade-off. Method 0 has the fastest encoding and method 6 has the best quality and larger encoding overhead. We select $\textit{method}=6$ to show the best compression performance and select a range of quality factors to evaluate its performance.
    
    \item \textbf{Progressive JPEG:} Progressive JPEG is a progressive variant of JPEG that supports image decoding with partial image data.  We explore the quality factor empirically and report the results with $q=30$, which balances classification accuracy and compression ratio.

    \item\textbf{RNN-TFLite:} We include a recurrent neural network (RNN) method ~\cite{toderici2017full} as a state-of-the-art method for progressive neural compression. The model is trained with the same set of training images as our model. We benchmark the encoding latency of RNN TFLite, a quantized version of the original RNN model in TensorFlow Lite.
    With an image of size $224\times 224$, this method generates 784 bytes of data during each iteration. 
    Note that the original work proposes multiple network implementations. We adopt a representative one with long short-term memory (LSTM) as the recurrent unit and ``additive reconstruction'' as the image reconstruction approach from our decoder's output. We select the implementation without entropy coding to mitigate the complexity of the model. Regardless, as shown in Section~\ref{sec:experiment:overhead}, RNN methods remain prohibitively expensive for embedded devices.
    
    \item \textbf{Starfish~\cite{10.1145/3384419.3430769}:} Starfish is a state-of-the-art AE-based approach designed to tolerate data loss. During training, Starfish utilizes a dropout layer located at the bottleneck of the AE to make the system resilient to data loss. Since the original Starfish was evaluated on different datasets, we retrained Starfish with our training dataset and optimized the model with quantization and Huffman coding. Note that, to reduce AE complexity and encoding latency, Starfish evenly divide images into four patches and encode them individually.
    
    

\end{itemize}

\subsection{Encoding Overhead}
\label{sec:experiment:overhead}
For each image compression method, we measure the image encoding overhead on the embedded device (Raspberry Pi 4). 
As shown in Table~\ref{tab:encoding_latency}, {\PNC} takes on average 9.8ms to encode the image into features of sorted importance. Additional ``feature quantization'' and ``Huffman encoding'' take an additional 2.0ms. 
The average encoding overhead is hence $9.8+2.0=11.8$ms\footnote{The average encoding overhead can be reduced to $11.4$ms if two threads are used to run the TFLite inference for the encoder. However, since the JPEG and WebP implementations in the Python image library, Pillow, used in our experiments, only support single-threaded encoding/decoding, for fairness of comparison, we also kept {\PNC} single-threaded in our experiments.}. WebP encodes an image in 6.0ms with $\textit{method}=0$, but the overhead grows significantly to over 32.5ms for $\textit{method}=6$, much longer than {\PNC}. 
Progressive JPEG incurs an average overhead of 5.8ms. 

In contrast, the neural compression baselines have much higher overheads than {\PNC} and the traditional compression methods. Starfish takes an AE with larger numbers of intermediate layers and channels than {\PNC}, resulting in a higher encoding overhead of 62.9ms for only the first image patch (out of a total of four), \textit{51.1ms} longer than PNC. The TFLite-optimized RNN baseline (RNN-TFLite) takes on average 1.9s to encode an image for one iteration when executing with 4 threads and takes 3.0s with only 1 thread. Hence, the prohibitively high complexity of RNN models makes them unsuitable for embedded devices. 

\begin{table}[t!]
\centering
\caption{Comparison of encoding overhead. For different WebP methods, $m$ represents the \emph{method}. For all applicable methods, $q$ represents the quality factor.}
\label{tab:encoding_latency}
\begin{tabular}{ccc}
\hline
\hline
Method     & Configuration & Average Encoding Overhead (ms) \\
\hline
\multirow{2}{*}{PNC}         & \#thread $=1$ & 11.8                   \\
           & \#thread $=2$ & 11.4                   \\
\multirow{4}{*}{WebP}       & $m=0,q=0\ $     & 6.0                    \\
           & $m=0,q=20$    & 6.7                    \\
           & $m=6,q=0\ $     & 32.5                   \\
           & $m=6,q=20$    & 48.6                   \\
Prog. JPEG & $q=30$        & 5.8                    \\
Starfish   & One Patch     & 62.9                   \\
RNN-TFLite & One Iteration &  1900    \\
\hline\hline
\end{tabular}
\end{table}

\subsection{Performance under Different Deadlines and Bandwidth} 
\label{sec:experiment:rt_perf}

\subsubsection{Experimental Setups}
\label{sec:experiment:deployment}

\begin{figure}[t!]
    \centering
    \includegraphics[width=0.9\linewidth]{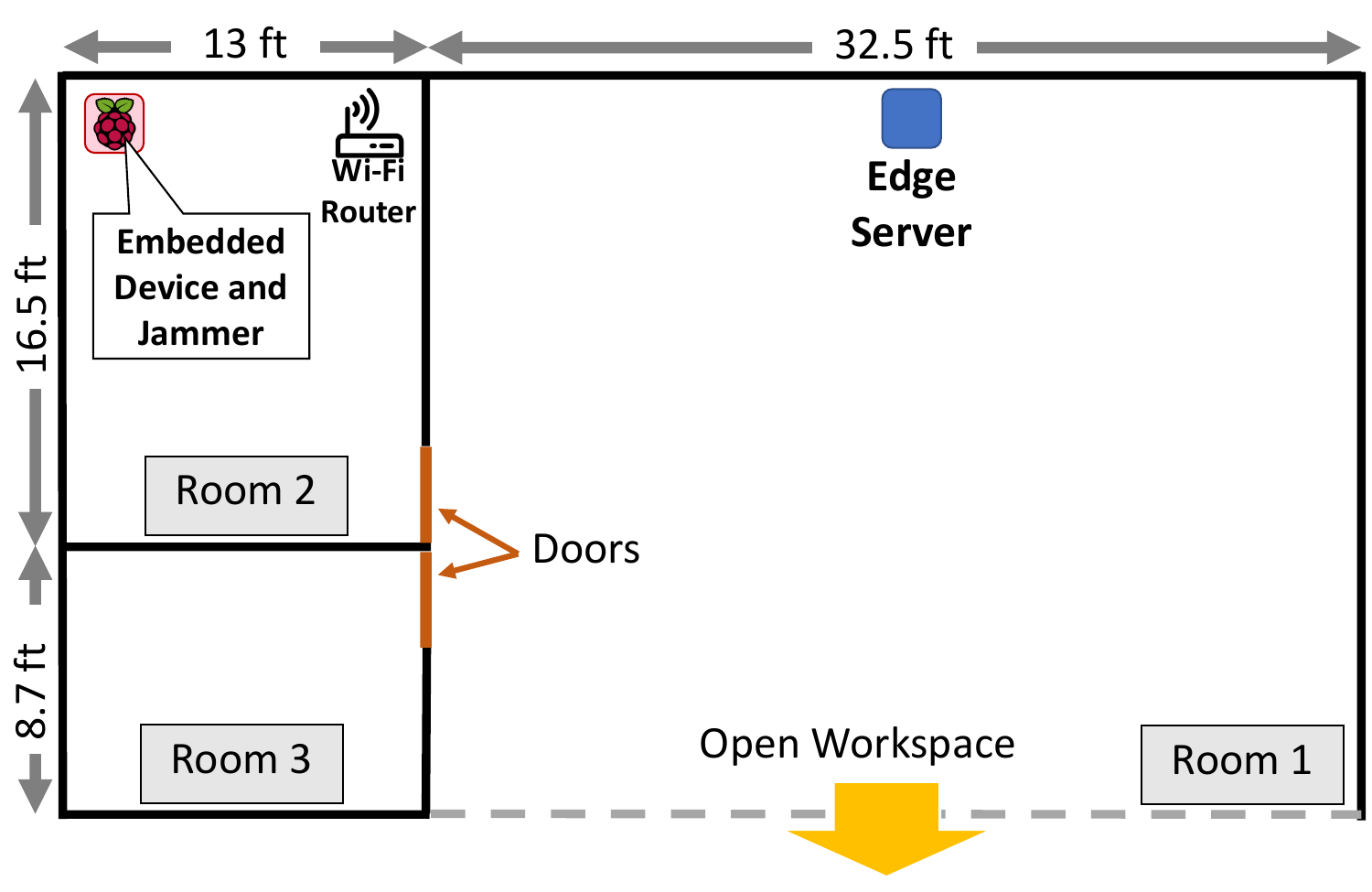}
    \caption{Layout of the indoor space used for the experiments. To minimize interferences from other environmental factors, all  experiments are conducted at night when no one is present.}
    \label{fig:floor_plan}
\end{figure}

We deploy our image classification system in a multi-room indoor space as shown in Fig.~\ref{fig:floor_plan}. The system comprises the embedded device, the jammer, and the edge server described in Section~\ref{sec:implementation:hw}. The server and the device are placed in Room 1 and Room 2, respectively, as shown in Fig.~\ref{fig:floor_plan}. The device offloads encoded images to the server over an IEEE 802.15.4 wireless link. The jammer (a Raspberry Pi 3 Model B) is placed close to the client and generates traffic at different rates using \texttt{iperf3}.  It communicates with a Wi-Fi router on a 2.4GHz Wi-Fi channel that overlaps with the IEEE 802.15.4 channel used for image offloading. Jamming intensity is varied
by controlling \texttt{iperf3}'s transmission rate. 


We conduct the experiments under (1) different network scenarios and (2) different timing constraints. We create three network scenarios with different levels of wireless interference and, consequently, different bandwidths between the device and the server:
\begin{itemize}
    \item \textbf{Scenario 1 (No Jamming):} There is no jamming traffic.
    \item \textbf{Scenario 2 (Light Jamming):} \texttt{iperf3} on the jammer generates traffic at a rate of 10 Mbps.
    \item \textbf{Scenario 3 (Heavy Jamming):}  \texttt{iperf3} on the jammer generates traffic at a rate of 20 Mbps.
\end{itemize}
For each network scenario, we repeat the experiments with different image arrival periods ($T$) of 300, 500, and 700 ms  (and hence image offloading deadlines of approximately the same values).






\begin{figure*}
  \begin{subfigure}{0.29\textwidth}
    \includegraphics[width=\linewidth]{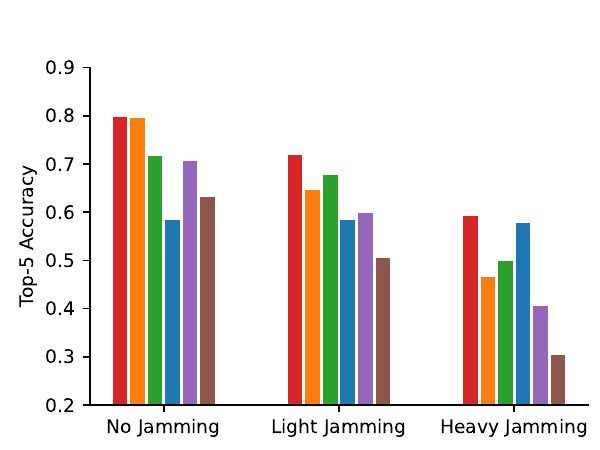}
    \caption{$T$ = 300ms}
    \label{fig:exp:experimental_results_bar:subfigure1}
  \end{subfigure}%
  \hfill
  \begin{subfigure}{0.29\textwidth}
    \includegraphics[width=\linewidth]{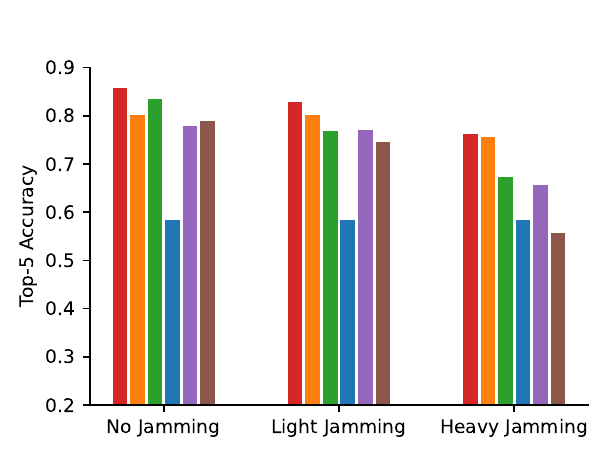}
    \caption{$T$ = 500ms}
    \label{fig:exp:experimental_results_bar:subfigure2}
  \end{subfigure}%
  \hfill
  \begin{subfigure}{0.29\textwidth}
    \includegraphics[width=\linewidth]{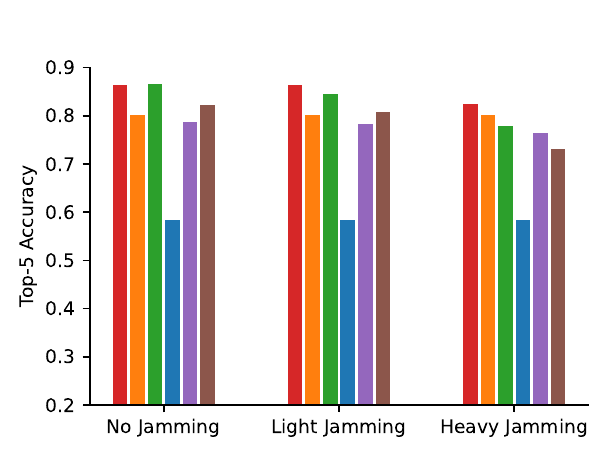}
    \caption{$T$ = 700ms}
    \label{fig:exp:experimental_results_bar:subfigure3}
  \end{subfigure}
  \begin{subfigure}{0.1\textwidth}
    \includegraphics[width=\linewidth]{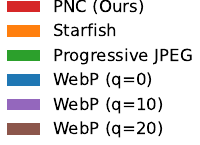}
    \vspace{2\baselineskip}
    \label{fig:exp:experimental_results_bar:legend_1}
  \end{subfigure}
  \caption{Classification accuracy. Each subfigure shows the results in a network scenario with different image arrival periods.} 
  \label{fig:exp:experimental_results_bar}
\end{figure*}


\begin{figure*}
\minipage{0.885\textwidth}
  \begin{subfigure}{0.32\linewidth}
    \includegraphics[width=\linewidth]{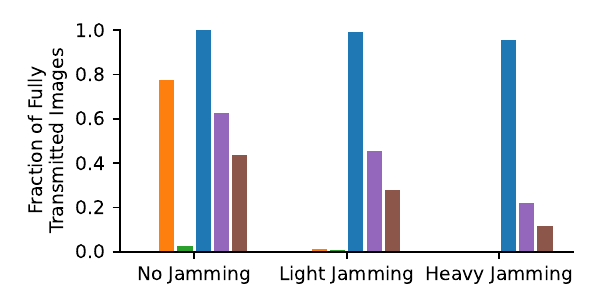}
    \caption{$T$ = 300ms}
    \label{fig:exp:detailed_analysis:subfigure1}
  \end{subfigure}%
  \hfill
  \begin{subfigure}{0.32\linewidth}
    \includegraphics[width=\linewidth]{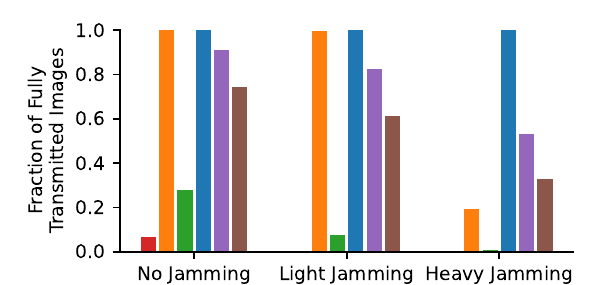}
    \caption{$T$ = 500ms}
    \label{fig:exp:detailed_analysis:subfigure3}
  \end{subfigure}%
  \hfill
  \begin{subfigure}{0.32\linewidth}
    \includegraphics[width=\linewidth]{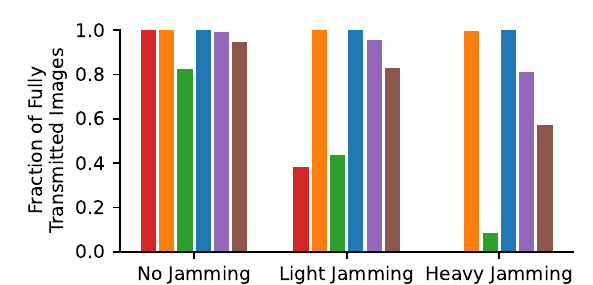}
    \caption{$T$ = 700ms}
    \label{fig:exp:detailed_analysis:subfigure5}
  \end{subfigure}
\endminipage
\hfill
\minipage{0.1\textwidth}
    \begin{subfigure}{1\textwidth}
        \includegraphics[width=\linewidth]{figures/plot_exp_acc_legend.pdf}
        \label{fig:exp:experimental_results_bar:legend_2}
    
    \end{subfigure}
\endminipage

\caption{Fraction of images completely offloaded within the time limit. Each subfigure shows the results in a network scenario with different image arrival periods.}
\label{fig:exp:fully_offloaded}
\end{figure*}

\begin{figure*}
\minipage{0.885\textwidth}  
  \begin{subfigure}{0.32\linewidth}
    \includegraphics[width=\linewidth]{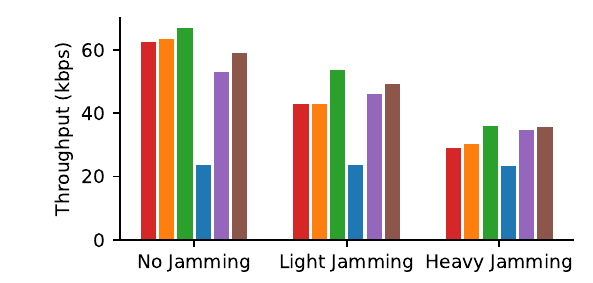}
    \caption{$T$ = 300ms}
    \label{fig:exp:detailed_analysis:subfigure2}
  \end{subfigure}%
  \hfill
  \begin{subfigure}{0.32\linewidth}
    \includegraphics[width=\linewidth]{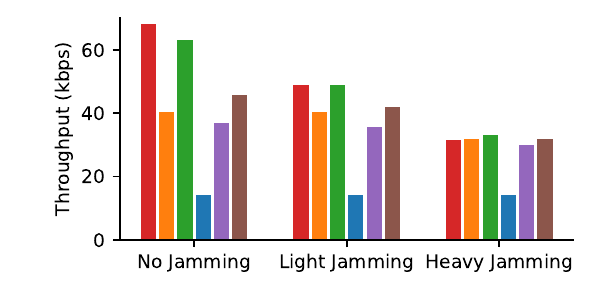}
    \caption{$T$ = 500ms}
    \label{fig:exp:detailed_analysis:subfigure4}
  \end{subfigure}
  \hfill
  \begin{subfigure}{0.32\linewidth}
    \includegraphics[width=\linewidth]{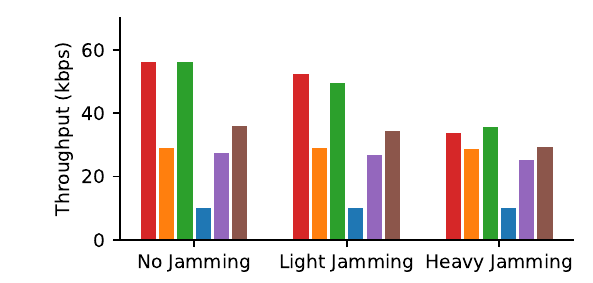}
    \caption{$T$ = 700ms}
    \label{fig:exp:detailed_analysis:subfigure6}
  \end{subfigure}
\endminipage
\hfill
\minipage{0.1\textwidth}
\begin{subfigure}{1\textwidth}
    \includegraphics[width=\linewidth]{figures/plot_exp_acc_legend.pdf}
    \label{fig:exp:experimental_results_bar:legend_3}
  \end{subfigure}
\endminipage
  \caption{Network throughput during image offloading. Each subfigure shows the results in a network scenario with different image arrival periods.}
  \label{fig:exp:throughput}
\end{figure*}

We conduct experiments on our testbed with a fixed set of 2,000 images drawn from the ImageNet validation set. As our system terminates the offloading of an image upon the arrival of the first feature of the next image, all the timing constraints for image offloading are met in the experiments. Henceforth, we evaluate the classification performance of {\PNC} and the baselines under different network scenarios and timing constraints. We use a standard metric, \emph{top-$n$ accuracy}, which is commonly used in previous edge computing works~\cite{chakrabarti2020real, yao2020deep, qiu2022adaptive}, to quantify classification performance.  Specifically, as per Eq.~\eqref{eqn:problem_formulation:top_n_acc} for $n=5$, top-5 accuracy assigns a score (reward) of 1 for the received image if its ground-truth class is among the 5 most likely classes produced by the classifier and 0 otherwise. 

Fig.~\ref{fig:exp:experimental_results_bar} shows the top-5 accuracy of {\PNC} and the baselines including the non-progressive baseline, WebP ($\textit{method}=6$), the progressive baseline, Progressive JPEG (with the best configuration, $q=30$), and the neural-compression baseline, Starfish. 
We leave RNN out due to its prohibitive encoding overhead on embedded devices (see Section~\ref{sec:experiment:overhead}). To further analyze the behavior of different solutions, we also show the fractions of images fully offloaded within the time constraints (Fig. 7) and the network throughput during image offloading (Fig. 8). 

\subsubsection{Classification Performance}

As expected, the image arrival period ($T$) and the available network bandwidth both affect the amount of data that can be transmitted to the server and, therefore, the methods' classification accuracy, \ie lower accuracy when $T$ decreases, or network jamming worsens. 

We first look at the performance of \textbf{WebP} with different quality levels ($q=0, 10, 20$). As a traditional compression approach not optimized for inference, WebP generally delivered poor classification performance. Moreover, as a non-progressive approach, WebP cannot adapt to different deadlines and network conditions. For instance, WebP with the \textit{highest} quality ($q=20$) achieved the highest classification accuracy among the WebP variants with \textit{T}=700ms and no jamming (Fig. 6c). In contrast, WebP with the \textit{lowest} quality ($q=0$) achieved the highest classification accuracy among WebP variants with  \textit{T}=300ms and heavy jamming (Fig. 6a). This is consistent with the fact that WebP ($q=20$) is able to offload most of the high-quality images fully in the former case, as allowed by the longer deadline and high bandwidth. In contrast, WebP ($q=20$) can offload only about 10\% of the images fully in the latter case. On the other hand, while WebP ($q=0$) is able to fully offload almost all the images in all the testing scenarios, its low quality limits the classification accuracy. This is consistent with the low throughput of WebP ($q=0$) in all the experiments (Fig. 8), which shows WebP ($q=0$) is not able to utilize the bandwidth available due to its aggressive compression at the cost of quality. We note that all variants of WebP generally achieve low throughput in comparison to {\PNC} and the other baselines due to the small data sizes produced by WebP encoding.

We then analyze the progressive baseline, \textbf{Progressive JPEG}. In contrast to WebP, Progressive JPEG can effectively adapt to deadlines and network conditions. As shown in Fig. 6, with Progressive JPEG, classification accuracy improves with longer deadlines and less network interference, which allows Progressive JPEG to offload more data. This is consistent with Fig. 8, which shows the system achieves higher throughout with Progressive JPEG in the presence of lower interference. Interestingly, the system achieves its classification performance even though only small fractions of the images are fully offloaded with Progressive JPEG (Fig. 7), thanks to its progressive encoding. However, Progressive JPEG consistently underperforms compared to {\PNC}, which suggests the strength of neural compression optimized for inference. The only scenario when Progressive JPEG and {\PNC} achieve similar classification accuracy is when \textit{T}=700ms with no jamming. Given the long deadline and large bandwidth in this scenario, {\PNC} successfully transmits all the features associated with each image by the deadline, while Progressive JPEG manages to refine many of the images to the point where the reconstructed image is sufficient for accurate image classification (Fig. 8).


The neural compression baseline, \textbf{Starfish}, is designed to tolerate data loss or incomplete data. As a result, it delivers relatively robust performance across different scenarios. However, Starfish has two limitations. First, it does not differentiate features when generating and transmitting them. When the deadline and bandwidth are severely limited, e.g., \textit{T}=300ms with heavy jamming, it suffers a significant loss in accuracy as the features offloaded are not sufficient for classification. In contrast, through its tail-drop strategy, {\PNC} generates features with different importance for classification and sends the features in order of importance. As a result, {\PNC} manages to offload the most important features within the timing constraint, even in the presence of heavy jamming. This ensures that {\PNC} consistently outperforms Starfish including in scenarios with tight timing constraints and low bandwidth. In addition, Starfish cannot take advantage of situations when a larger amount of data are allowed to be transmitted to the server, \ie with \textit{T}=700ms. This is because the Starfish encoder produces a relatively small amount of data (on average 2.53KB) after compression, 
which in turn upper-bounds the amount of data that can be offloaded. This is consistent in the fact Starfish offloads most of the images fully with \textit{T}=700ms (Fig. 7c) while showing low network throughput (Fig. 8c) because it does not generate enough data to fully utilize the bandwidth available.

\textbf{{\PNC}} consistently outperforms all the baselines across all the scenarios in our experiments. Like Progressive JPEG, {\PNC} maintains robust classification performance (Fig. 6) while offloading only small fractions of images fully to the server (Fig. 8). The consistent performance is attributed to its novel strategy of progressively transmitting the most important features first. Furthermore, it is able to increase throughput by exploiting available bandwidth when there is less inference (Fig. 8).

\begin{figure}[t]
    \centering
    \includegraphics[width=1\linewidth]{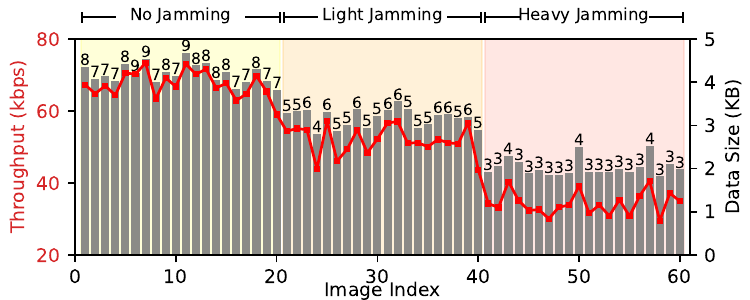}
    \caption{A typical run of the offloading process under varying network bandwidth. The red curve shows the throughput of offloading traffic. The gray bars represent the data size of the fully offloaded features. The black number above each bar represents the number of features fully offloaded to the edge server. 
    Note that Huffman encoding encodes each feature into unequal sizes so that the number of features and data size are not always proportional to each other.}
    \label{fig:exp:adapt}
\end{figure}

\begin{figure}[t]
    \centering
    \includegraphics[width=0.85\linewidth]{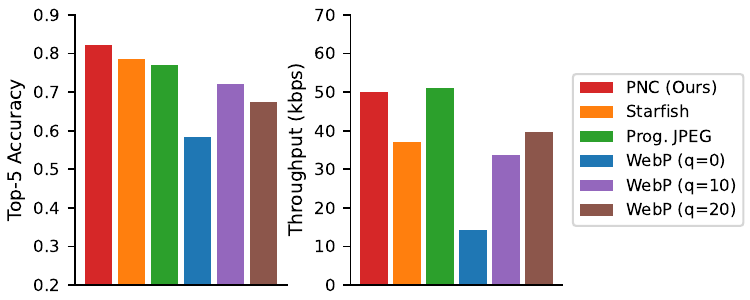}
    \caption{Image classification accuracy and offloading throughput under varying scenarios.}
    \label{fig:exp:dynamic_exp}
\end{figure}

\subsection{Performance in Changing Scenarios}
\label{sec:experiment:varying}
This subsection further examines {\PNC}'s ability to adapt to bandwidth variations by switching among different network scenarios dynamically during each experiment. Fig.~\ref{fig:exp:adapt} shows a consecutive sequence of 60 offloaded images with an image arrival period of $T$ = 500ms and decreasing network bandwidth. We use \texttt{iperf3} to create varying levels of interference
according to the No, Light, and Heavy Jamming network scenarios in sequence, each lasting for 10 seconds. As the level of interference increases, the throughput of {\PNC} drops along with the data size and the number of features offloaded to the server, which illustrates {\PNC} adapting to the variations in network bandwidth by delivering a different number of features for classification.

We then conduct a set of experiments under varying bandwidth, where we switch between ``No Jamming'' and ``Heavy Jamming'' every 10s to create significant changes to available bandwidth. The performance of {\PNC} and the baselines are shown in Fig.~\ref{fig:exp:dynamic_exp}. Consistent with Fig.~\ref{fig:exp:experimental_results_bar} and the corresponding analysis, {\PNC} achieves the best classification performance, followed by Starfish and Progressive JPEG. This confirms that {\PNC} can maintain high levels of classification accuracy by prioritizing important features in the presence of low bandwidth.  Like {\PNC}, Progressive JPEG also fully utilizes bandwidth but its encoding is optimized for reconstruction quality instead of classification accuracy. Starfish underutilizes the network and has a lower throughput, especially when no network jamming exists, but still outperforms Progressive JPEG in this experiment due to its advantage in neural encoding. The WebP baselines again perform the worst because of the impact of partially received images in the presence of heavy network jamming.

In summary, the experiments demonstrate that {\PNC} (1) runs efficiently on resource-constrained IoT or embedded devices, (2) achieves superior classification accuracy in comparison to both standard compression methods and state-of-the-art neural compression approaches under bandwidth and deadline constraints, and (3) can adapt to bandwidth variations while maintaining high levels of classification accuracy. 

\section{Conclusion}

In this paper, we proposed a progressive neural compression framework, {\PNC}, for offloading images under timing constraints for edge-assisted classification. In contrast to neural compression solutions that usually encode inputs into fixed sizes, {\PNC} is inherently designed to support progressive encoding optimized for image classification. As a result, {\PNC} can classify images at high accuracy even when the encoded image is partially received, thereby allowing the system to adapt to different timing constraints and unpredictable bandwidth fluctuations often experienced on wireless sensor networks. Empirical evaluation of {\PNC} on a testbed demonstrates its superior classification performance over state-of-the-art compression solutions under varying wireless conditions and timing constraints, as well as its high encoding efficiency on IoT devices. In the future, the idea of progressive neural compression may be extended to other real-time applications, \ie automatic speech recognition and object detection, that involve compression with DNNs.



\section*{Acknowledgment}
This research was supported by the Fullgraf Foundation and in part by NSF Grant CNS-2006530.

\bibliographystyle{IEEEtran}
\bibliography{reference}

\end{document}